\documentclass[a4paper,fleqn,usenatbib]{mnras}

\usepackage{newtxtext,newtxmath}

\usepackage[T1]{fontenc}
\usepackage{ae,aecompl}

\usepackage{graphicx}	
\usepackage{amsmath}	
\usepackage{amssymb}	

\title[Activity in a solar RV timeseries]{Reducing activity-induced variations in a radial-velocity timeseries of the Sun as a star}

\author[A. F. Lanza et al.]{
A. F. Lanza,$^{1}$\thanks{E-mail: antonino.lanza@inaf.it (AFL)}
A. Collier Cameron,$^{2}$
R. D. Haywood$^{3}\thanks{NASA Sagan Fellow}$
\\
$^{1}$INAF-Osservatorio Astrofisico di Catania, Via S.~Sofia,78 - 95123 Catania, Italy\\
$^{2}$Centre for Exoplanet Science, SUPA, School of Physics and Astronomy, University of St Andrews, St Andrews, KY16 9SS, UK\\
$^{3}$Harvard-Smithsonian Center for Astrophysics, 60 Garden Street, Cambridge, MA 02138, USA\\
}

\date{Accepted XXX. Received YYY; in original form ZZZ}

\pubyear{2019}

\begin{document}
\label{firstpage}
\pagerange{\pageref{firstpage}--\pageref{lastpage}}
\maketitle

\begin{abstract}
The radial velocity of the Sun as a star is affected by its surface convection and magnetic activity.  The moments of the cross-correlation function between the solar spectrum and a binary line mask contain information about the stellar radial velocity and line-profile distortions caused by stellar activity. As additional  indicators, we consider the disc-averaged magnetic flux and the filling factor of the magnetic regions. Here we show that the activity-induced radial-velocity fluctuations are reduced when we apply a kernel regression to these activity indicators. The disc-averaged magnetic flux  proves to be the best activity proxy over a timescale of one month and gives a standard deviation of the regression residuals of $1.04$ m/s,  more than a factor of 2.8 smaller than the standard deviation of the original radial velocity fluctuations. This result has been achieved thanks to the high-cadence and time continuity of the observations that simultaneously sample both the radial velocity and the activity proxies. 
\end{abstract}

\begin{keywords}
techniques: radial velocities -- sun: activity -- planets and satellites: detection 
\end{keywords}



\section{Introduction}
The discovery and characterization of Earth-like planets is an active area of modern astronomy and requires precise radial velocity (hereafter RV) measurements to confirm transit candidates and evaluate their mass. The best speedometers currently available, such as HARPS and HARPS-N, allow us to reach a precision better than 1~m/s on bright targets \citep[e.g.,][]{Pepeetal11} that is comparable with the RV signal induced by an Earth-mass planet in the habitable zone of an M-type dwarf star.  An even higher precision of the order of 0.1~m/s will be reached by the new spectrograph ESPRESSO at the VLT telescopes \citep[][]{Pepeetal14}. At those levels of precision, stellar phenomena such as p-mode oscillations, convection, and magnetic activity contribute to produce apparent RV variations in late-type stars with similar or larger amplitude than the searched planetary signals  \citep[][]{Dumusqueetal11a,Dumusqueetal11b,Fischeretal16}. 

The Sun can be used as a template to study the RV variations induced by stellar activity because we can resolve its surface observing the effects of different brightness and convective inhomogeneities. To this purpose, \citet{Haywoodetal16} observed the Sun as a star by measuring the RV variations of its light reflected by the asteroid 4/Vesta and compared the observations with a model reconstruction based on images and magnetograms obtained by the instruments on board of the Solar Dynamic Observatory (SDO). Their dataset covers a couple of months, while a similar work by \citet{Lanzaetal16}, using the Moon, the Galileian satellites and several asteroids as reflectors, covers most of an eleven-year activity cycle, although with a rather sparse sampling. On the other hand, \citet{Dumusqueetal15} and \citet{Phillipsetal16} used an integrating sphere to obtain the spectrum of the Sun as a star and measure its RV variations by means of HARPS-N. Their time series now covers almost three years during the declining phase of solar cycle 24 \citep{Milbourneetal19}.

In the case of distant stars, it is not possible to derive resolved maps of their surface inhomogeneities comparable to those of the Sun. Therefore, it is better to look for proxies of magnetic activity that can be extracted from the same spectrum used to measure the RV of the stars themselves. Specifically, the cross-correlation function (hereafter CCF) between the spectrum and the binary line mask used to measure the stellar RV \citep{Baranneetal96} can be particularly useful for this purpose because it has a signal-to-noise ratio of the order of $10^{3}$ that allows to detect the tiny line profile distortions induced by small surface features such as those found in stars with a low activity level that are the primary targets for Earth-like planet searches. 

Recently, \citet{Lanzaetal18} proposed to use indicators derived from the CCF in combination with a kernel regression (hereafter KR) to reduce the activity-induced RV variations  in a sample of 15 sun-like stars. They were selected to have slow rotation  ($v \sin i \leq 5$~km/s) and low activity with a chromospheric  index $\log R^{\prime}_{\rm HK} \leq -4.95$,   comparable with that of the Sun at the minimum of its eleven-year cycle.  The bisector inverse span (hereafter BIS), the full width at half maximum (FWHM) and a new indicator called $V_{\rm asy(mod)}$ proved to have the best performances giving significant regressions with the RV variations in $\sim 50$ percent of the stars and reducing the standard deviation of the RV by approximately a factor of two down to $1.1-1.5$ ~m/s. 

Another proxy for stellar activity is the disc-integrated unsigned magnetic flux that can be measured from the Zeeman excess broadening of spectral lines with a Land\'e factor greater than the unity without the need for spectropolarimetric measurements \citep[e.g.,][]{Reiners12}. This proxy is particularly important because it is directly related to the magnetic field, that is the physical agent responsible for stellar activity. 

In the present work, we extend the approach of \citet{Lanzaetal18} to the Sun as a star using the observations of \citet{Haywoodetal16} and demonstrate the advantage of the KR technique, finding regression residuals down to $\sim 1.0$~m/s on a  timescale of one-month, that corresponds to about one solar rotation. 

The main advantage of the KR approach is its conceptual simplicity. More sophisticated state-of-the-art techniques model the activity-induced component of the RV variation by means of Gaussian Processes \citep{Haywoodetal14,Grunblattetal15}. The most recent applications of such techniques include a simultaneous multivariate modelling of the variations of the RV and activity proxies  by means of an underlying Gaussian Process \citep{Rajpauletal15,Jonesetal17}. These models can include not only the correlation between the RV variations and the activity proxies, but also  the time derivatives of the proxies themselves that gives the possibility to model phase shifts and time delays between the different quantities. The price to pay for this greater  flexibility is a greater model complexity. On the other hand, KR is much simpler and can be used for computing a preliminary  regression between the RV and different activity proxies to provide a quick evaluation of their relative performances. 

\section{Observations}
\label{observations}
\citet{Haywoodetal16} monitored the RV variations of the Sun as a star by observing the sunlight reflected by the asteroid 4/Vesta from the 29th September 2011 to the 7th December 2011 for a total of 98 datapoints acquired over  37 nights during a period of rather high activity in solar cycle 24. The spectra were acquired with HARPS at the 3.6-m telescope at La Silla with a precision of $0.75 \pm 0.25$~m/s. The RV variations were corrected for the relativistic Doppler shifts induced by the motion of Vesta and of the observer and reduced to the barycentre of the Sun \citep[see][for details]{Haywoodetal16}. These data are plotted in the top panel of Fig.~2 and are listed in Table~A.1 of \citet{Haywoodetal16}, while the files with the CCFs can be downloaded from the URL indicated in the same paper. 

The surface of Vesta is not homogeneous and the asteroid is rotating with a period of  5.34~hr that induce a RV modulation in the timeseries as discussed in Sect.~2.4.1 of \citet{Haywoodetal16}. To correct for this effect, we subtract the modulation due to Vesta axial rotation as derived in Sect.~4.1 of that paper, i.e., $\Delta RV_{\rm Vesta}(t) = C \cos [2\pi-\lambda(t)] + S \sin [2\pi - \lambda(t)]$, where the coefficients $C=2.19$ and $S=0.55$~m/s, and $\lambda$ is the apparent planetographic longitude of Vesta at the flux-weighted mid-times $t$ of the HARPS observations \citep[see Table~A.1 in][]{Haywoodetal16}. 

The radial velocity of the Sun, after the correction for the rotation of Vesta, is plotted in the middle panel of Fig.~2 of \citet{Haywoodetal16}. The datapoints show a higher dispersion during the first part of the observations  because the finite diameter of the disc of Vesta in combination with small guiding errors during the exposures produced an imperfect sampling of the regions of the asteroid disc having different rotational velocities.  This rotational imbalance in the spectrum of the light falling inside the fibre entrance was responsible for the fluctuations in the RV of Vesta. During the second half of the observational campaign, this problem was avoided thanks  to the smaller apparent diameter of the asteroid that was at a larger distance from the Earth. 

\section{Methods}
\label{methods}
We follow the same approach as in \citet{Lanzaetal18} to whom we refer the reader for details. First we compute the CCF profile indicators, that is the contrast (relative depth of the CCF at its central wavelength), FWHM, BIS \citep[cf.][]{Quelozetal01}, $\Delta V$ of  \citet{Nardettoetal06}, and $V_{\rm asy (mod)}$ \citep[see Sect.~3.1 of][for their definitions]{Lanzaetal18}. While the BIS and $\Delta V$ measure the shape of the bisector and the asymmetry of the red and blue wings of the CCF, respectively, $V_{\rm asy(mod)}$ is associated with the variations of the slopes of  the red and blue wings of the CCF that are more directly related to the variations of the RV measurements. This happens because the RV is derived by fitting the CCF with a Gaussian, thus the parts of the CCF with the highest slope and larger number of photoelectrons play a more relevant role  in determining the central RV of the Gaussian fit. 

For completeness, we add to the above indicators the chromospheric index $\log R^{\prime}_{\rm HK}$, derived from the Ca II H\&K line profiles of the HARPS spectra,  the disc-averaged unsigned magnetic flux $| \hat{B}_{\rm obs} | $, and  the filling factor $f$ of  the magnetic regions over the solar disc, derived from the SDO images and magnetograms. They are listed in Table A.1 of \citet{Haywoodetal16}. 

We consider the regressions between the above activity indicators and the RV variation of the Sun as a star. Since the relationships between RV and activity indicators are neither linear nor monotonic \citep[see, for example, Fig.~6 in][showing the relationship between the RV and the BIS]{Boisseetal11}, we apply a kernel regression (KR) to model them. The basic principle of KR is to compute a regression between the RV data  and one of the activity indicators giving different weights to the datapoints. Specifically, we give the highest weights to the points that are closer in time and in the value of the indicator to a given point and compute as many regressions as the datapoints to investigate the correlation between the RV, the time, and the considered indicator. The weight of each datapoint is specified by the kernel function that is tuned by varying two parameters, called bandwidths (see below). We apply a locally linear regression to fit the RV at the time  $t_{k}$  with a value of the indicator $x_{k}$, where $k=1, ..., N$ is the index of the point and $N$ the total number of datapoints.  We compute the regression by minimizing the function $Z$ with respect to the coefficients $\beta_{0}$ and $\beta_{1}$, where $Z$ is given by: 
\begin{equation}
Z = \sum_{i=1}^{N} [RV(t_{i}) - \beta_{0} -\beta_{1}(x_{i} - x_{k})]^{2} W(x_{i}-x_{k}, t_{i} - t_{k}),
\end{equation}
and the kernel is:
\begin{equation}
W(x_{i}-x_{k}, t_{i} - t_{k}) = \exp \left\{ - \left[ \left( \frac{x_{i}-x_{k}}{h_{x}} \right)^{2} + \left( \frac{t_{i}-t_{k}}{h_{t}} \right)^{2} \right] \right\},
\end{equation}
where $h_{x}$ and $h_{t}$ are the bandwidths. The method to compute the KR and the optimal bandwidths is described in Sect.~3 of \citet{Lanzaetal18} to which we refer the reader. 

In the present work, we use only symmetrical kernels to allow a direct comparison with the results of \citet{Lanzaetal18}, but a kernel that is asymmetric in time can effectively account for a time lag between the RV and a proxy indicator. Such a lag is responsible for the hysteresis observed between the RV and the CCF profile indicators \citep[cf. ][]{Boisseetal11,Figueiraetal13}. Moreover, a lag between the RV and the chromospheric index $\log R^{\prime}_{\rm HK}$ has been noted in GJ~674 \citep{Bonfilsetal07}, GJ~176 \citep{Forveilleetal09}, and
HD~41248 \citep{Santosetal14}. 

One drawback of the KR is the possibility of overfitting the data. The occurrence of overfitting can be tested by looking at the distribution of the residuals because it is expected to be remarkably non-Gaussian in that case with an overabundance of small residuals with respect to the frequency expected from a normal distribution.  

When the datapoints and the residuals are normally distributed, it is possible to compute analytically the significance $p$ of the KR by means of the Fischer-Snedecor statistics $F$ based on the ratio of the $\chi^{2}$ of the original data to that of the KR residuals. The effective number of degrees of freedom of the regression $\nu$ can be evaluated analytically from the data \citep[cf.][]{Lanzaetal18}.  The value of $p$ is the probability of attaining the given value of the $F$ statistics in the case of purely random fluctuations. As such, it cannot be used to compare the relative performances of the KRs with different indicators, but only to reject the null hypothesis that there is no correlation between the predictor variable (i.e., the indicator) and the RVs. For simplicity, to compare the different indicators, we use the standard deviation of their KR residuals as our figure of merit to rank them (see Sect.~\ref{results}). Note that this is not based on any statistics argument, but it serves our purpose of reducing the intrinsic RV variations as much as possible.  

\section{Results}
\label{results}
\begin{figure*}
\centerline{
\includegraphics[height=20cm,width=10cm,angle=270]{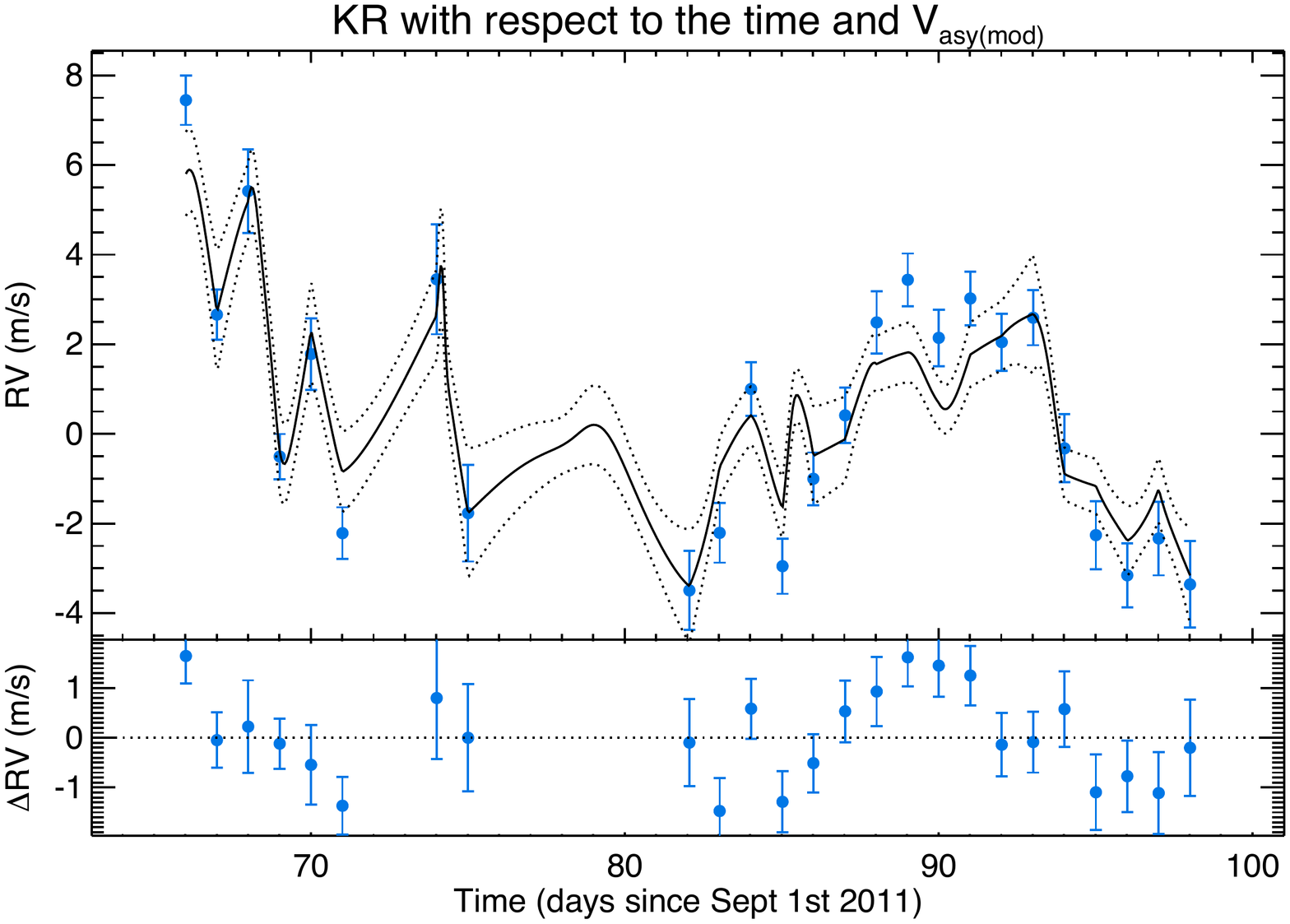}} 
\caption{Top panel: the second part of the RV timeseries of the Sun as a star of \citet{Haywoodetal16} (blue filled circles) and the interpolated KR with respect to the time and the CCF asymmetry indicator $V_{\rm asy (mod)}$ (solid line). The dotted lines indicate the $\pm \, \sigma$ intervals of the interpolated KR as computed with the method described in Sect.~3.2 of \citet{Lanzaetal18}. Lower panel: the timeseries of the residuals of the uninterpolated KR  (blue filled circles). }
\label{KR_Vasy_second_part}
\end{figure*}
\begin{figure*}
\centerline{
\includegraphics[height=20cm,width=10cm,angle=270]{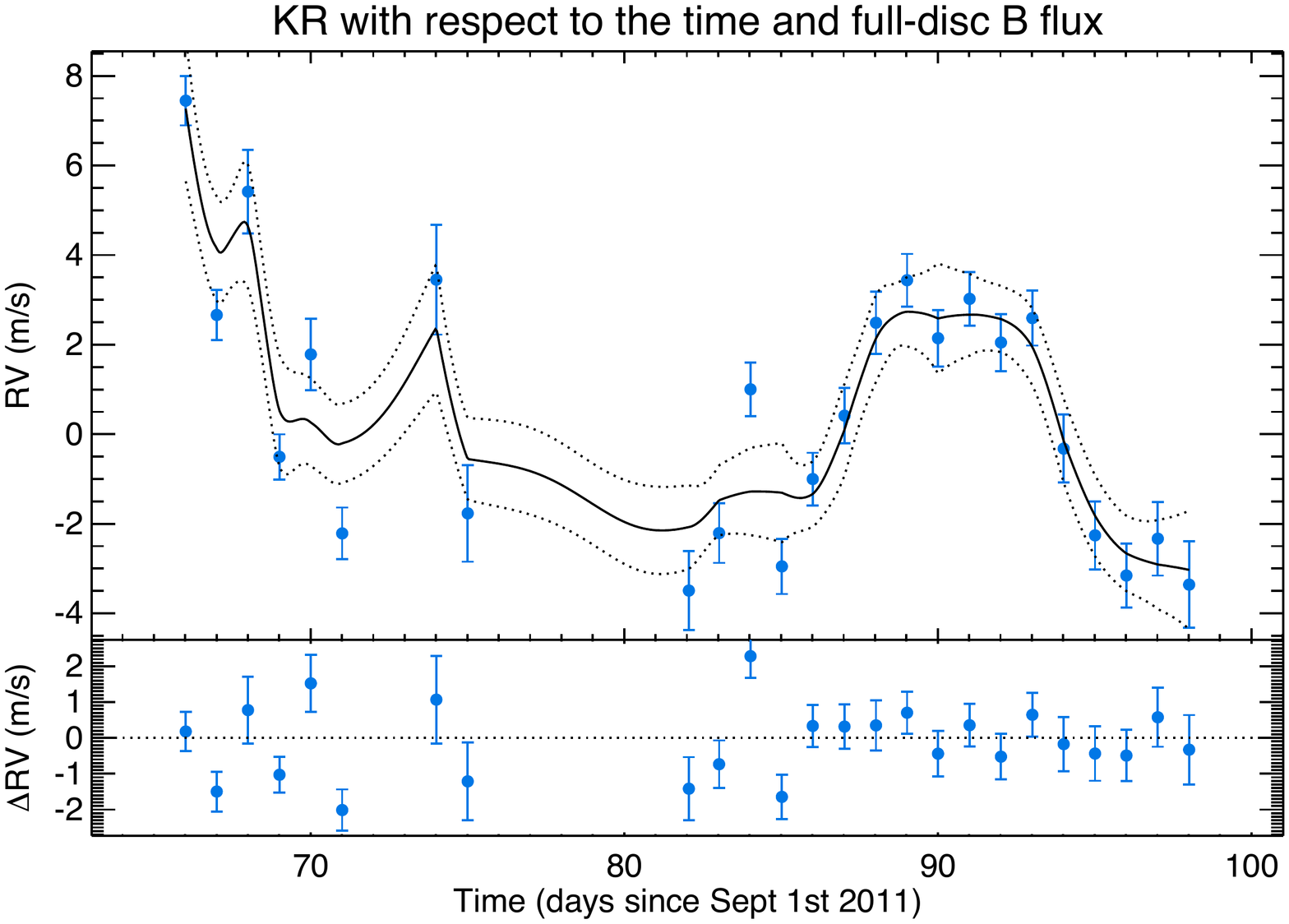}} 
\caption{Same as Fig.~\ref{KR_Vasy_second_part}, but for the mean unsigned magnetic field $|\hat{B}_{\rm obs}|$. }
\label{KR_Bfield_second_part}
\end{figure*}
\begin{figure*}
\centerline{
\includegraphics[height=20cm,width=10cm,angle=270]{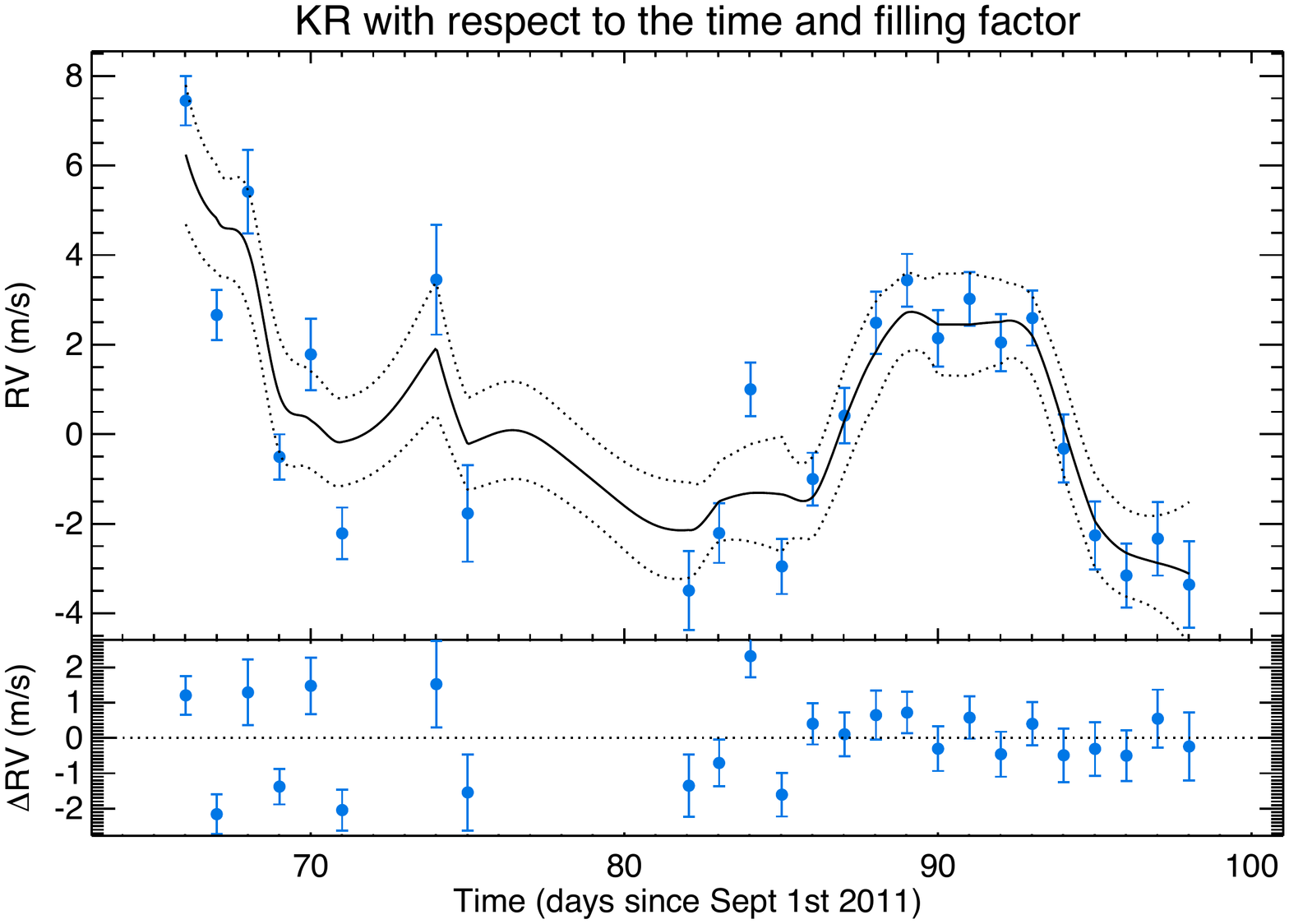}} 
\caption{Same as Fig.~\ref{KR_Vasy_second_part}, but for the filling factor $f$ of the magnetic field. }
\label{KR_filling_second_part}
\end{figure*}
\begin{figure*}
\centerline{
\includegraphics[height=20cm,width=10cm,angle=270]{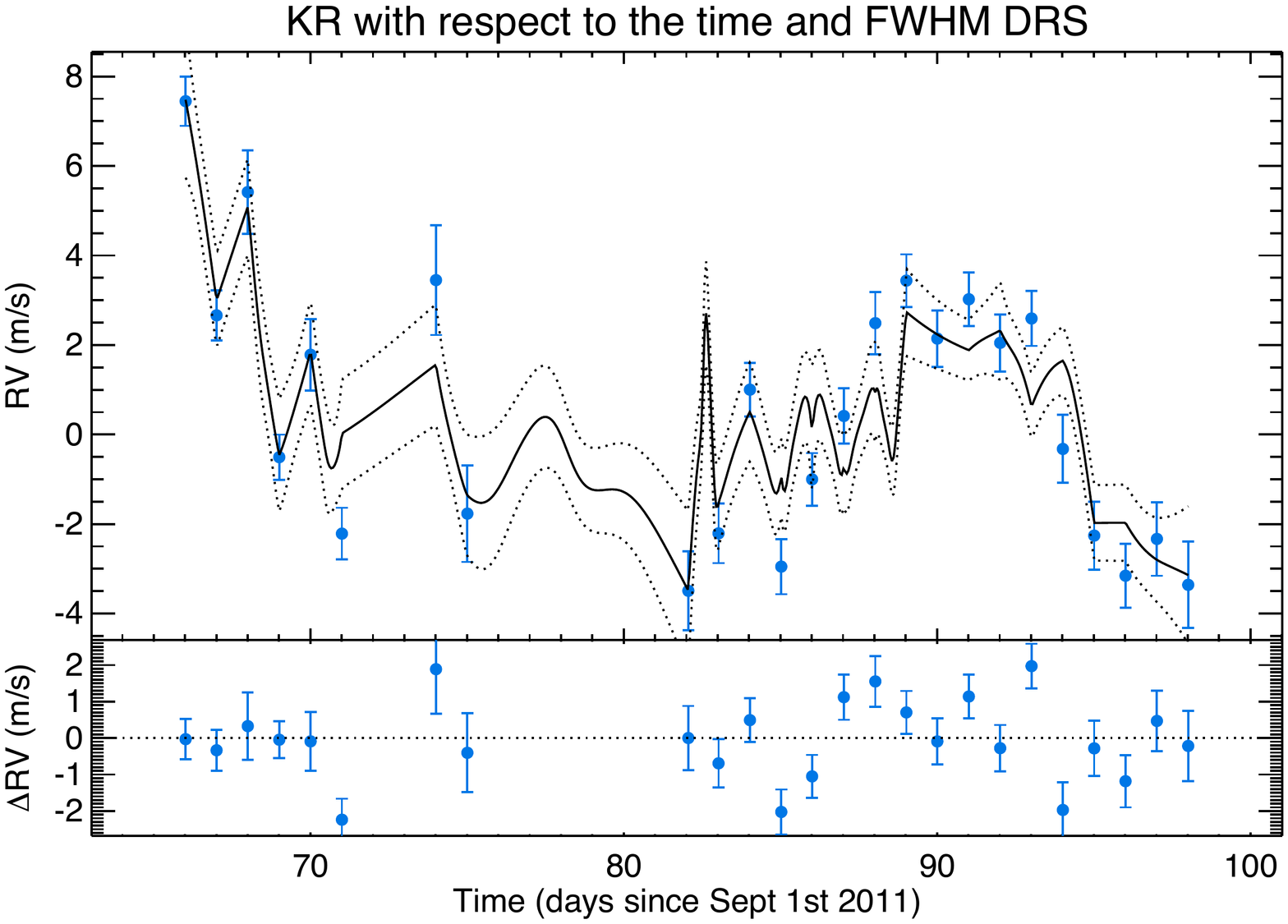}} 
\caption{Same as Fig.~\ref{KR_Vasy_second_part}, but for the FWHM of the CCF  as given by the Data Reduction System (DRS) of HARPS. }
\label{KR_FWHM_second_part}
\end{figure*}

We restrict our analysis to the second part of the RV time series of \citet{Haywoodetal16}, that is starting from Julian Date 2455870.5142, because of the  systematic errors present in the first part of the dataset as discussed in Sect.~\ref{observations}. To reduce the effect of photospheric convective motions, we averaged the RV measurements taken during the same night as suggested by \citet{Dumusqueetal11a}. From the individual 76 measurements, this leads to $N=25$ datapoints with a standard deviation of 2.969~m/s. 

We reject the null hypotheses  (p-value $< 0.01$) in all the kernel regressions reported in Table~\ref{Table2} where we list,  from the left to the right, the activity indicator, the standard deviation $\sigma$ of the 25 datapoints, the standard deviation $\sigma_{\rm KR}$ of the residuals after application of the KR, the bandwidths $h_{t}$ and $h_{x}$, the Fischer-Snedecor statistics $F$, the $p$-value giving the significance, the effective number $\nu$ of degrees of freedom of the KR, and the probability $P_{\rm KS}$ that the distribution of the KR residuals comes from a normal distribution with zero mean and standard deviation of 0.75~m/s, evaluated with the Kolmogorov-Smirnov test \citep{Pressetal02}.  

The different KRs are listed in order of increasing standard deviation $\sigma_{\rm KR}$ of their residuals. We checked that all the distributions of the residuals are indistinguishable from a normal distribution as indicated by the relatively large values of $P_{\rm KS}$ assuming a threshold p-value of 0.01.  As indicated in Sect.~\ref{methods}, we  choose the standard deviation of the residuals $\sigma_{\rm KR}$ as our figure of merit to rank our KRs in order to reduce the RV variability induced by stellar activity as much as possible. 

We plot the KR with respect to the time and our best indicators $V_{\rm asy (mod)}$, $|\hat{B}_{\rm obs}|$, $f$, and FWHM of the CCF, in the top panels of Figs.~\ref{KR_Vasy_second_part}, \ref{KR_Bfield_second_part}, \ref{KR_filling_second_part}, and~\ref{KR_FWHM_second_part}, while the residuals of the regressions are in the lower panels, respectively.  We do not plot the KRs with respect to $\Delta V$, BIS, $\log R^{\prime}_{\rm HK}$, and contrast because of  their larger $\sigma_{\rm KR}$. The KRs plotted in the top panels of those figures have been interpolated over an evenly sampled time grid using the method in Sect.~3.2 of \citet{Lanzaetal18}, while the residuals in the lower panels are those of the uninterpolated KRs (note also the different RV-axis scales in the two panels). 
A gap in the time series produces some modulation and a slightly larger uncertainty in the KR because the slope of the local regression is unconstrained due to the lack of datapoints,  while the KR tries to match the regressions with different slopes coming from the two sides of the gaps. 

The KRs with respect to $V_{\rm asy (mod)}$ and FWHM show more oscillations than those with respect to $|\hat{B}_{\rm obs}|$ and $f$.  In the case of $V_{\rm asy (mod)}$ in Fig.~\ref{KR_Vasy_second_part}, before day $\sim 75$, the regression tries to fit the individual datapoints rather than providing a smoother function to reproduce the overall variation, thus showing some overfitting. In the subsequent intervals,  the regression  is systematically below the datapoints between days $88-91$ and systematically above during days $95-98$.  Similar drawbacks are noted for the KR with respect to the FWHM in Fig.~\ref{KR_FWHM_second_part} with the additional effect of a sharp variation in the slope of the regression at day 82.5 where two local linear regressions with largely different slopes come to match with each other. 

In conclusions, looking at the individual KRs, we obtain relevant information,  in addition to the criterion of minimizing $\sigma_{\rm KR}$, that allows us to select the best KR. Considering the smoothness of the regression and the minimum $\sigma_{\rm KR}$, we select the KR with respect to the time and the mean magnetic field $|\hat{B}_{\rm obs}|$ as the best one.    For this regression, the effective number of degrees of freedom $\nu$ is smaller by a factor of 2.5 than the number of datapoints indicating a meaningful regression \citep[cf. ][]{Lanzaetal18}.  

\begin{figure*}
\centerline{
\includegraphics[height=20cm,width=10cm,angle=270]{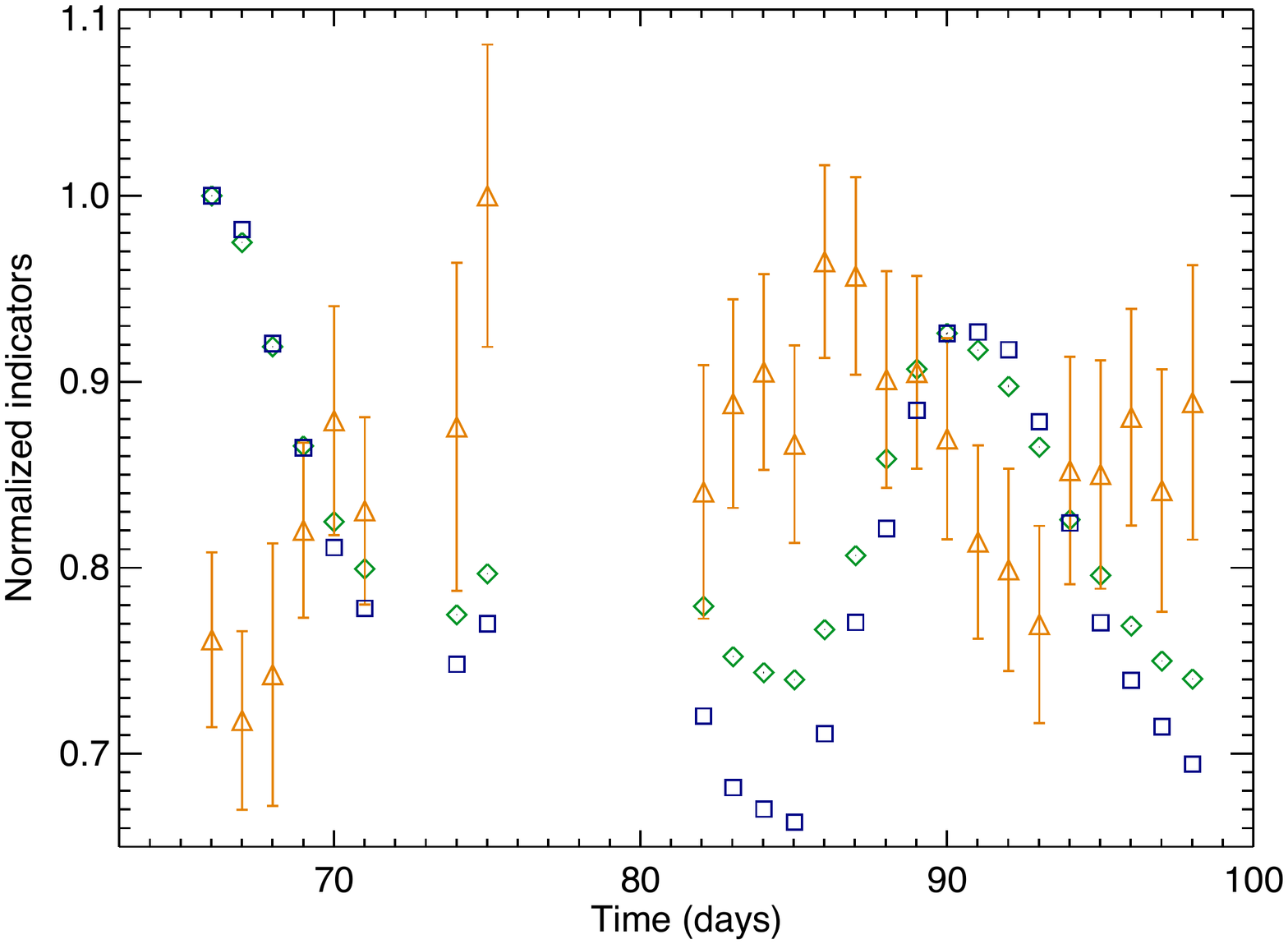}} 
\caption{The CCF asymmetry indicator $V_{\rm asy (mod)}$ (open orange triangles), the mean magnetic field $|\hat{B}_{\rm obs}|$ (open green diamonds), and the filling factor $f$ of magnetic areas (open blue squares) vs.  the time in our time series. All the indicators have been normalized to their maximum values, respectively. The values of the magnetic proxies come from \citet{Haywoodetal16} and their errorbars are much shorter than the size of the symbols. Note the close association between the two magnetic indicators $|\hat{B}_{\rm obs}|$ and $f$, while the variation of the CCF asymmetry index $V_{\rm asy (mod)}$ is  in antiphase with respect to those magnetic indexes.}
\label{indicators_plots}
\end{figure*}

In principle, the time bandwidth $h_{t}$ may provide an evolutionary timescale for the active regions  responsible for the RV variations. As a matter of fact, a KR can adjust its time bandwidth in an attempt to reproduce short-term variability the activity indicator is not capable of  accounting for. This could be related to physical processes, such as supergranular convection, that are not directly affecting any of our considered activity proxies. A similar phenomenon has been noted with other regression techniques applied to reduce the impact of stellar activity such as Gaussian Processes \citep[e.g., Sect.~2.3 of ][]{RasmussenWilliams06}. However, in our specific case, we see that the value of $h_{\rm t} \simeq 4.6$~days is similar to the typical timescale during which the variations of our best indicators and of the RV have approximately constant slopes as can be seen by comparing Figs.~\ref{KR_Bfield_second_part} and~\ref{indicators_plots}.     A longer time series is required to draw conclusions on the association between the value of $h_{t}$ and the properties of the main active regions dominating the RV variations. The very recent release of the first extended dataset from the HARPS-N solar telescope \citep{Milbourneetal19} will allow future investigations to clarify this issue. 
\begin{table*}
	\centering
	\caption{Kernel regression parameters and statistics for the second part of the nightly binned time series in \citet{Haywoodetal16}. The units of measure of the values of $h_{x}$ are the same as those of the corresponding indicators, that is, gauss for $|\hat{B}_{\rm obs}|$, km/s for FWHM and BIS, and percent for $f$, while the remaining indicators are non-dimensional. }
	\label{Table2}
	\begin{tabular}{ccccccccc} 
		\hline
		Indicator & $\sigma$ & $\sigma_{\rm KR}$ & $h_{t}$ & $h_{x}$ & $F$ & $p$ & $\nu$  & $P_{\rm KS}$ \\
		  & (m/s) & (m/s) & (d) & & & &  \\
		\hline
	$V_{\rm asy(mod)}$  &    2.969    &     0.932    &     4.560 &    1.534e-03    &     14.12243   &   0.000014   &     10.032 &  0.778\\
	$|\hat{B}_{\rm obs}|$     &    2.969    &     1.037   &      4.560  &   6.743e-01  &   12.40123  &    0.000026    &     9.374 &  0.508\\
        FWHM  &    2.969     &    1.116     &    4.560   &  1.593e-03       &    10.29986   &   0.000080   &      9.522 &  0.676 \\
         $f$  &     2.969     &    1.185   &      4.560  &   4.653e-01  &    8.56638  &    0.000237   &      9.743 &  0.219 \\
         	 $\Delta V$       &   2.969      &   1.193    &     4.560  &   4.612e-03    &   11.82464   &   0.000029 &         8.020 & 0.251\\
        BIS      &   2.969      &   1.267     &    4.560   &  9.666e-04       &   8.20888   &   0.000277    &     9.129 & 0.039 \\        
        $\log R^{\prime}_{\rm HK}$   &   2.969    &     1.491     &    4.560  &   7.999e-03       &   4.95419  &    0.003885    &     9.578 & 0.190\\       
        Contrast   &    2.969    &     1.554    &     4.560 &    1.582e-04   &    4.04510  &    0.009892   &     10.103 & 0.081\\        
		            \hline
	\end{tabular}
\end{table*}
\section{Discussion and conclusions}
 The application of KR to a one-month timeseries of the RV of the Sun as a star shows that the disc-averaged magnetic flux $|\hat{B}_{\rm obs}|$ is the best proxy of activity effects  and can be used to reduce the standard deviation of the RV residuals down to 1.04 m/s, that is by a factor of 2.86 with respect to the standard deviation of the original RV data. This level of noise reduction is approaching the value of $\sim 4$ that \citet{Halletal18} deem necessary to detect an Earth-mass planet on a yearly orbit. 

The conclusion that the disc-averaged magnetic flux is the best proxy to correct the RV variations due to activity is in line with  \citet{Haywoodetal16} who considered a model based on the resolved maps of the solar disc as obtained by the instruments on board of SDO. It is also comparable with what can be achieved by applying Gaussian Processes to model activity in other stars as in, e.g.,  \citet{Haywoodetal14,Haywoodetal18}. 

The performance of the activity proxies based on the FWHM or the asymmetry of the CCF is lower than that of the disc-averaged unsigned magnetic flux and more  similar to the results obtained by modelling Sun-like stars in \citet{Lanzaetal18}  where KR  gave a reduction of the standard deviation of the activity-induced RV variations by a factor of $\approx 2$. In the application to the present solar dataset, we see that the KR with respect to these CCF indicators can in some cases suffer from excessive smoothing or in others from overfitting the observed RV variations. This is the case of the indicator $V_{\rm asy(mod)}$ that, although achieving the best performance in terms of reduction of the standard deviation of the residuals, shows clear signs of both these effects. 

It is important to note that the present performance is related to the exquisite time sampling of the present RV measurements that allowed us to average out the variations on a timescales of a few hours associated with photospheric convection as well as  to follow the modulations produced by stellar activity with very good continuity. These requirements of frequent sampling and continuity in time  have been pointed out also by, e.g., \citet{Anglada-Escudeetal16}, \citet{Lopez-Moralesetal16}, \citet{Haywoodetal18} who applied different techniques including Gaussian Processes. 

\section*{Acknowledgements}
The authors are grateful to the referee, Dr. Rodrigo Fernando D\'{\i}az, for a careful reading of their manuscript and several valuable comments that helped them to improve their work. 
AFL, ACC, and RDH are grateful to the organizers of the international workshop "Observing the Sun as a star: would we find the solar system if we saw it\,?" at the University of G\"ottigen in September 2018 during which the original idea of this work was conceived.  AFL acknowledges support by INAF/Frontiera through the "Progetti Premiali" funding scheme of the Italian Ministry of Education, University, and Research. ACC acknowledges support from the Science \& Technology Facilities Council (STFC) consolidated grant number ST/R000824/1. This work was performed in part under contract with the California Institute of Technology (Caltech)/Jet Propulsion Laboratory (JPL) funded by NASA through the Sagan Fellowship Program executed by the NASA Exoplanet Science Institute (R.D.H.).
%
%
%




\bsp	
\label{lastpage}
\end{document}